\documentclass[twocolumn]{article}
\usepackage{graphicx}

\makeatletter
\newenvironment{mybibliography}
 {\let\@afterindentfalse\@afterindenttrue
  \setlength{\leftskip}{\parindent}
  \setlength{\parindent}{-\leftskip}}
\makeatother

\begin{document}

\title{A Word Communication System with Caregiver Assist for Amyotrophic Lateral Sclerosis Patients in Completely and Almost Completely Locked-in State}

\author{Kuniaki Ozawa$^1$\thanks{Corresponding author. E-mail: ozawa@m.ieice.org},
        Masayoshi Naito$^1$,
        Naoki Tanaka$^{1,2}$,
        and Shiryu Wada$^3$  \\[6pt]
        $^1$\textit{\normalsize Research Institute of Industrial Technology, Toyo University, Kawagoe, Japan}  \\
        $^2$\textit{\normalsize Department of Biomedical Engineering, Toyo University, Kawagoe, Japan}         \\
        $^3$\textit{\normalsize Double Research and Development Co., Ltd., Zama, Japan}
}

\date{}

\maketitle

\begin{abstract}
People with heavy physical impairment such as amyotrophic lateral sclerosis (ALS) in a completely locked-in state (CLIS) suffer from inability to express their thoughts to others. To solve this problem, many brain-computer interface (BCI) systems have been developed, but they have not proven sufficient for CLIS. In this paper, we propose a word communication system: a BCI with caregiver assist, in which caregivers play an active role in helping patients express a word. We report here that four ALS patients in almost CLIS and one in CLIS succeeded in expressing their own words (in Japanese) in response to wh-questions that could not be answered ``yes/no.'' Each subject selected vowels (maximum three) contained in the word that he or she wanted to express in a sequential way, by using a ``yes/no'' communication aid based on near-infrared light. Then, a caregiver entered the selected vowels into a dictionary with vowel entries, which returned candidate words having those vowels. When there were no appropriate words, the caregiver changed one vowel and searched again or started over from the beginning. When an appropriate word was selected, it was confirmed by the subject via ``yes/no'' answers. Three subjects expressed ``yes'' for the selected word at least six times out of eight (reliability of 91.0\% by a statistical measure), one subject (in CLIS) did so five times out of eight (74.6\%), and one subject three times out of four (81.3\%). We have thus taken the first step toward a practical word communication system for such patients.
\end{abstract}

\vspace{5mm}

\noindent
{\small 
\textbf{Keywords: communication, yes/no, word expression, NIRS, ALS, CLIS, BCI, caregiver assist}
}

\section{Introduction}
Communication is essential for anyone to fully live life. Some people with amyotrophic lateral sclerosis (ALS), however, cannot control their muscles when the disease progresses to an extreme called a completely locked-in state (CLIS). Hence, they cannot even express their ``yes/no'' intention using ordinary means. Besides, from the viewpoint of communication, such people’s answers should not be limited to ``yes/no'' (Wolpaw and Wolpaw, 2012).  Accordingly, brain-computer interface (BCI) research has sought to enable spelling of words since 1999 (Birbaumer et al., 1999). Systems using P300 signals such as BCI2000 (Schalk and Mellinger, 2010) have been proposed. It is usually impossible, however, for CLIS patients to see a letter matrix because of their drooping eyelids.

Therefore, spelling systems using auditory stimuli have been researched (Simon et al., 2015; Halder et al., 2016). Those works showed the possibility for healthy subjects to spell words with that approach, but they did not prove successful in providing ALS patients in CLIS with the means to spell a word. Instead of using P300 signals, BCIs based on near-infrared spectroscopy (NIRS) have been investigated (Sereshkeh et al., 2019, Hong et al. 2020). NIRS-based BCIs,  proposed first in 2004 (Coyle et al., 2004; Sitaram et al. 2007), achieved ``yes/no'' communication for ALS patients in CLIS (Naito et al., 2007; Gallegos-Ayala et al., 2014; Chaudhary et al., 2017). Still, it is difficult for such patients to spell a word even with a NIRS-based BCI. The reason is that the patients are expected to perform too many steps to correctly choose the row and column (for the consonant and vowel, respectively, in Japanese) in a letter matrix, which tends to result in choosing wrong letters and burdening them with correcting misspellings.

Therefore, we propose a practical online word communication system for home use that consists of a ``yes/no'' communication aid using near-infrared light and a special dictionary with vowel entries. The system only requires a patient to choose three vowels and then answer ``yes/no'' to candidate words including those vowels. To share the burden of spelling words, caregiver assist is essential for this system. Specifically, it requires a caregiver to select an appropriate word in the dictionary, to change one vowel when an appropriate word is not found in the dictionary, and to start over from choosing three vowels when the patient denies an appropriate word.

In this paper, we report experimental results of using this system with four ALS subjects in almost CLIS (on the verge of CLIS) and one subject in CLIS. By cooperating with their caregivers, the subjects could successfully express words in response to wh-questions such as ``What is your favorite animal?''

\section{Method}
\label{method}
The study was approved by the Toyo University Ethical Review Board for Medical and Health Research Involving Human Subjects and the Institutional Review Board of the Public Health Research Foundation. To avoid tiring the subject, the time of one experimental session was limited to 30 minutes, with a maximum of 16 questions. Informed consent was obtained from the legal representatives in the subjects' families beforehand.
Written, informed consent was obtained from the patient' legal guardian for the publication of any potentially identifiable images or data included in this article.

\subsection{Word Communication}
\subsubsection{Outline of Proposed System}
\label{outline}

The Japanese kana syllabary consists of five vowels, A, I, U, E, and O; 14 consonants, K, S, T, N, H, M, Y, R, W, B, D, G, P, and Z; and a special letter ``NN'' without a vowel. Every Japanese kana character other than a vowel or NN consists of a consonant and a vowel. If a patient tries expressing a word consisting of 3 Japanese characters without any cooperation, for example, he or she needs to get correct yes/no classification as many as 25.5 times (8.5 times for each character: 3 plus 5.5 times in the Japanese kana syllabary of $5\times 10$ matrix), which will make it impossible to reach the true word.

\begin{figure*}
\centering
\includegraphics[width=153mm]{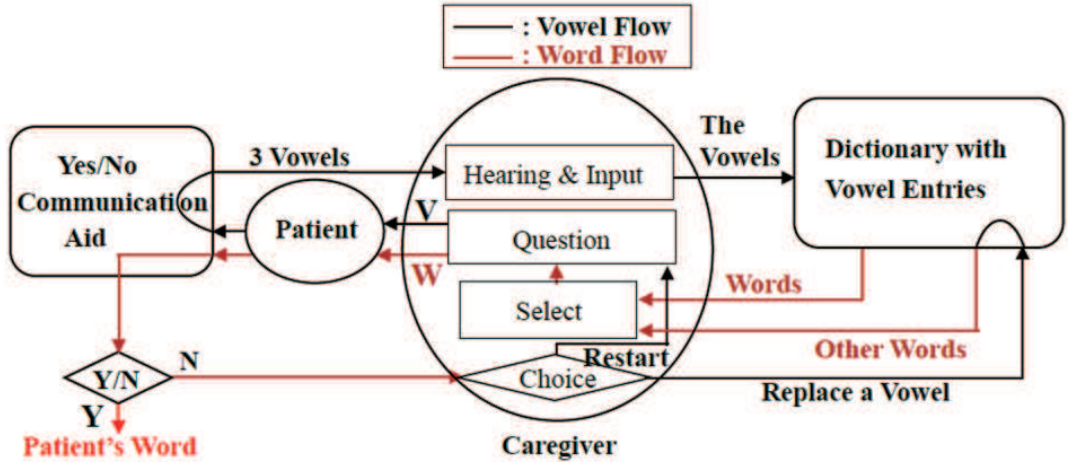}
\caption{\small Word communication system with caregiver assist. The caregiver plays three active roles: one is to ask the patient for three vowels, another is to select an appropriate word among words having those vowels that the dictionary returns, and the third is to check several times whether it is the right one. The caregiver accepts it if the patient confirms it enough times.}
\label{systemfig}
\end{figure*}

Figure \ref{systemfig} shows the framework of the proposed word communication system. It consists of a ``yes/no'' communication aid, a dictionary with vowel entries, a patient, and a caregiver. The communication aid is ``Shin Kokoro Gatari'' (``New Heart Teller,'' a product of Double Research and Development Co., Ltd, 2016), which is described in the next section. We prepared a new dictionary that returns words containing input vowels. The dictionary receives three vowels as input and returns the words containing them in the given order. For example, if a sequence ``E, I, A'' is input, then the dictionary returns 125 words, such as 
\textit{\textbf{\textbf{\underline{ei}}}g\textbf{\textbf{\underline{a}}}} (movie) and 
\textit{h\textbf{\underline{ei}}w\textbf{a}} (peace). It has about 10,000 words, according to the fact that Japanese adults usually use 10,000 words in daily life (Miyakoshi, 2010).

Using the communication aid, a caregiver asks a patient to give the first three vowels (via the scheme described below) contained in the word that the patient wants to express. Then, the caregiver consults the dictionary by entering the three vowels and obtains candidate words. The caregiver selects a word among them and asks the patient whether it is the right one using the communication aid. After repeating this process, if the patient denies all the selected words, then the caregiver either replaces a vowel with another one and consults the dictionary again or restarts from the beginning. Finally, the caregiver uses the communication aid to check several times whether the finally selected word is the right one, and he or she accepts the word if the patient confirms it enough times.

\begin{figure}
\centering
\includegraphics[width=80mm]{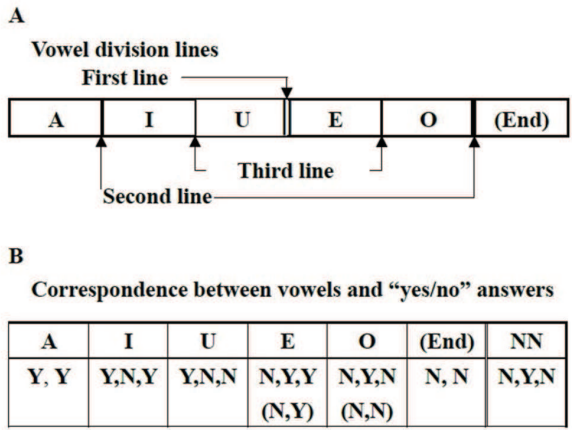}
\caption{\small Vowel table. (A) Vowel division lines. The lines divide the vowels into groups. The double line divides the vowels into \{A, I, U\} and \{E, O, (End)\}. The single lines divide the groups into \{A and I, U\} and \{E, O and (End)\}. Finally, the thin lines separate \{I and U\} and \{E and O\}. (B) Correspondence between vowels and ``yes/no'' answers. In the first vowel acquisition, because (End) can be excluded as the first choice, we assign ``no, yes'' to E and ``no, no'' to O. In the second or third vowel acquisition, when a patient answers ``no, yes, no,'' it is also assigned to the letter NN, which never occurs in word-initial position.}
\label{divisionlines}
\end{figure}

The scheme for determining a vowel works as follows. We use the dichotomic table shown in Fig. \ref{divisionlines}A. The patient is asked whether the vowel is in the left group and answers ``yes'' or ``no'' in a sequential way. For example, if the patient wants to select I, then he or she answers ``yes'' to the first question, ``Does your word contain A, I or U?'' (i.e., the group \{A $|$ I, U\} is selected from \{A, I, U $\|$ E, O\}). Then, the patient selects ``no'' (indicating the group \{I $|$ U\}), and then ``yes'' (indicating \{I\}). Figure \ref{divisionlines}B summarizes the correspondence of ``yes/no'' answers to the vowels.

\subsubsection{Confirmation of Expressed Words}
\label{conf}
To confirm that the selected word is the right one, we need a statistical measure. In setting the measure, we used Bayesian statistics with the binominal likelihood and uniform prior probability density. For the measure, we adopted a quantity obtained by inegrating the posterior probability density over the range $0.5<\theta \le 1$, where $\theta$ is the affirmative probability.

For confirmation, the caregiver uses the communication aid to ask the patient eight times whether the word is right or wrong. When seven answers are affirmative, the value of the measure is 0.98, i.e., 98.0\% promising. When six answers are affirmative, the value is 91.0\%. In these cases, the caregiver accepts the word as the right one. When five answers are affirmative, however, the value is 74.6\%. In that case, the caregiver is free to decide acceptance.

For one word, a pair of questions is asked four times. The pair consists of an affirmative form, ``Is it right that your word is `X'?'', and a negative form, ``Is it right that your word is not `X'?'' When a patient answers ``no'' to the negative-form question, the patient's intention is regarded as affirmative.

\subsection{Yes/No Communication Aid}
\label{aid}
\subsubsection{Measurement}
\label{Hbmes}
A patient is asked a question and changes his or her brain activity depending on whether the answer is ``yes'' or ``no.'' The changes in activity are measured with near-infrared light as changes in the prefrontal blood volume. The present aid has two probes, one on the left forehead and one on the right forehead. Each probe has a light source (LED, 840-nm wavelength) and a Si PIN photodetector located 30 mm from the light source. The sampling frequency is 10 Hz. The change in the intensity of measured light is transformed into logarithm. At 840 nm, the dynamics of oxy-hemoglobin is mainly detected.

Each measurement consists of three periods of 12 seconds each: resting, answering, and resting again. In the experiment reported here, the ``yes'' task in the answering period was mental arithmetic or fast mental singing, and the ``no'' task was slow mental humming or imagining a landscape, which was the same task as in the resting period.

\subsubsection{Method of Yes/No Classification}
\label{ynclassif}
The method of yes/no classification is an extension of that described in Naito et al. (Naito et al., 2007). There are two major differences: one is that we utilize the change in heart rate in addition to the change in the blood volume accompanied by brain activity, and the other is the use of a support vector machine (L1 SVM with a Gaussian kernel) as the classifier. In the SVM, we set the cost parameter $C$ to 1,000 and the variance $\sigma^2$ of the kernel to 30.

The change in blood volume accompanied by brain activity is obtained applying a low-pass filter with a cut-off frequency of 0.1 Hz. The change in heart rate is obtained from the pulse wave contained in the blood wave as follows. First, the pulse wave is extracted by a band-pass filter extracting components in the range of $f_p\pm 0.3$ Hz, where $f_p$ is the peak frequency searched in the range above 0.5 Hz. Second, the Hilbert transform is applied to the pulse wave to obtain the unwrapped phase $\phi_{pw}(t)$, from which the instantaneous heart rate is calculated as $\left(d\phi_{pw}(t)/dt\right)\left/ 2\pi \times 60\right.$. Finally, the heart rate is smoothed by the low-pass filter.

The vectors input to the SVM are two-dimensional. We prepare three sets of feature vectors from which the optimal one is to be selected: feature set 1 consists of the number of oscillations (oscillation number) of the changes in the heart rate and that of the blood volume; set 2 consists of the oscillation number and the maximum amplitude of the blood volume change; and set 3 consists of the oscillation number and the maximum amplitude of the heart rate change. We calculate these quantities by using the analytic signal obtained by applying the Hilbert transform to the data in a time window in which the data is used for analysis. The maximum amplitude is the maximum instantaneous amplitude of the analytic signal. The oscillation number is obtained by dividing the increment in the unwrapped instantaneous phase by $2\pi$. The maximum amplitude is rescaled so that its order of magnitude is the same as that of the oscillation number.

The optimal feature set and time window are selected automatically to give high performance of the SVM. The time window is searched every second in the region after the first resting period with the minimum width of 15 s. Because we found that the change in blood volume lags behind the change in heart rate by 3 s on average, we shift the time window for the blood volume back by 3 s.

The SVM is trained with three pairs of yes/no training data and a pair of test data for cross-validation. The SVM performance is checked for various choices of the feature set and the time window. The performance measure is the average of the classification accuracies on the training and test data. When more than one parameter set has the same performance measure result, we select the one giving the largest geometric margin in the SVM.

\subsection{Experiments}
\label{exp}
Word-expression experiments were performed once a week from November 24, 2018 to April 24, 2019 at the subjects' own homes. A subject's family chose themes and asked wh-questions, which cannot be answered with ``yes/no.'' As described in Section \ref{outline}, in response, a subject chose three vowels, each of which was selected via two or three ``yes/no'' answers, through the communication aid. The subject was informed of the answer after each ``yes/no'' selection. 
The subjects were also instructed that, when they failed to give the right ``yes/no'' answer, they were to just keep relaxing afterward in the vowel acquisition trial, because the right vowel could not be obtained regardless of whatever ``yes/no'' response they made anyway.

\section{Subjects}
\label{subjects}
Five ALS patients without communication ability were recruited, as listed in Table \ref{subjectswords} (Subjects A-E). The mean age was 59.8 years (SD of 19.6 years). Subject D was female and the others were male. Every subject used an artificial respirator. Subject C turned out to be in CLIS according to the diagnosis of an experienced neurologist. The other subjects were in a minimal communication state (MCS), which is   defined by ``severely reduced speed and delay in initiation of voluntary functions`` (Hayashi and Oppenheimer, 2003). Almost completely locked-in state is medically equivalent to MCS. 

All subjects and caregivers were trained once a week by expressing the subjects' birthplaces. It took four sessions (days), from November 7, 2018 to February 13, 2019, for the subjects to understand how to use the communication system.

\begin{table*}
\caption{Results of expressed words for each subject.}
\label{subjectswords}
\centering
\begin{tabular}{cclllc}

\hline
Subject  & Age (Sex)  &  \multicolumn{1}{c}{Wh-question}  &  \multicolumn{1}{c}{3 vowels}   &  \multicolumn{1}{c}{Expressed word}  &  Sessions  \\
\hline
A (MCS)  &  72 (M)  &  Favorite animal  & E, A, A &  \textit{M\textbf{\underline{e}}d\textbf{\underline{a}}k\textbf{\underline{a}}} (kilifish)  &  5  \\
B (MCS)	 &  51 (M)	& Comment on system	& U, A, O(NN) &	\textit{F\textbf{\underline{uann}}tei} (unstable)	&  3   \\
C (CLIS) &  30 (M)	& Favorite genre for reading  & E, I, I	& \textit{R\textbf{\underline{e}}k\textbf{\underline{i}}sh\textbf{\underline{i}}} (history) & 4  \\
D (MCS)	&   67 (F)	& Transportation method    & I, O, U  & \textit{H\textbf{\underline{i}}k\textbf{\underline{ou}}ki} (airplane)	   & 3  \\
 &  &  \ \ to return to hometown & &  &  \\
E (MCS)	&   79 (M)	& Comment on system  &  A, I, A   & \textit{\textbf{\underline{A}}r\textbf{\underline{i}}g\textbf{\underline{a}}tai} (thanks)  &  4   \\
\hline
\end{tabular}
\end{table*}

\section{Results}
\subsection{Expressed Words for Wh-Questions}
Table \ref{subjectswords} lists the wh-questions given by the caregivers and the words expressed by the subjects. The results were obtained from the first trial for each subject, and the caregivers had no know-how on selecting candidate words. Therefore, they had to take a trial-and-error approach. 

\subsection{Detailed Results}
\subsubsection{Subject A}
The theme ``favorite animal'' was given by his wife, who anticipated the answer ``dog'' or ``cat.'' His expressed word, however, was \textit{medaka} (killifish), which none of his caregivers expected. He took five days to express that word. It was confirmed by his affirmative answers six times out of eight.

On the first day, Subject A answered all ``no'' for the three vowels, which resulted in ``O, O, (End).'' Unfortunately, there were no appropriate candidate words in the dictionary. On the second day, his caregivers chose \textit{h\textbf{\underline{a}}t\textbf{\underline{o}}} (pigeon), \textit{n\textbf{\underline{e}}k\textbf{\underline{o}}} (cat), \textit{h\textbf{\underline{i}}t\textbf{\underline{o}}} (person), and ``other'' among 14 candidate animals in which one vowel differed from ``O, O.'' His wife asked him words that were divided into two groups of (1) \textit{hato} or \textit{neko} and (2) \textit{hito} or ``other.'' His answer was in group (1) three times out of four. Then, his wife was sure that his answer was \textit{neko} and asked him if it was right. He replied that it was wrong three times out of four. Again she asked about groups (1) and (2) and he answered ``no'' for both. The caregivers were at a loss and finished up for the day.

\begin{figure*}
\centering
\includegraphics[width=145mm]{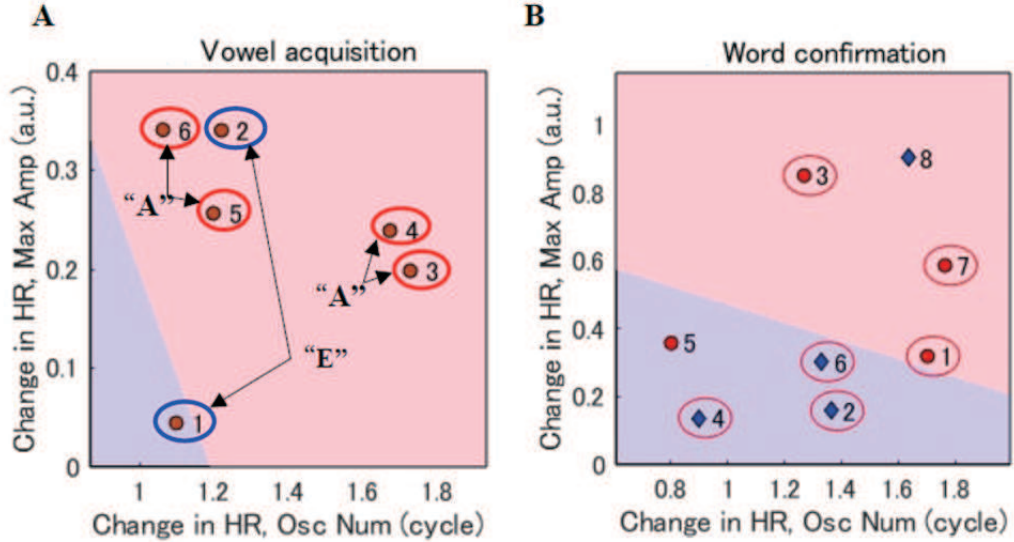}
\caption{\small Feature vectors for Subject A, with the pink region indicating ``yes'' and the light purple region indicating ``no''. (A) Feature vectors in vowel acquisition when the vowel sequence ``E, A, A'' was obtained, from the left forehead. The ``yes/no'' decision algorithm selected feature set 3 (HR denotes heart rate) and the time window from 14 to 29 s as optimal. (B) Feature vectors in word confirmation, from the right forehead. A red circular disc indicates a ``yes'' answer to the affirmative question, while a blue diamond indicates a ``no'' answer to the negative question. Thus, the red circular discs in the ``yes'' region (1, 3, 7) and the blue diamonds in the ``no'' region (2, 4, 6) are regarded as affirmative. The decision algorithm also selected feature set 3 in the confirmation session, but it selected the time window from 16 to 35 s.}
\label{vectors}
\end{figure*}

On the third day, the caregivers started from the beginning and asked Subject A for the three vowels. This time, he expressed ``E, A, A,'' as shown in Fig. \ref{vectors}A, for which there was one candidate word: \textit{m\textbf{\underline{e}}d\textbf{\underline{a}}k\textbf{\underline{a}}}. The caregivers tried to confirm it, and he answered ``yes'' only two times out of four, so they could not decide his word. On the fourth day, he expressed ``E, A, E.'' Though the caregivers found no candidate words in the dictionary, they noticed that ``E, A, \underline{E}'' differed by just one vowel from the previously expressed vowels ``E, A, \underline{A}.'' They asked him if ``E, A, A'' was right, and his answer was ``yes.'' Then, they asked if \textit{medaka} was right, to which he answered ``yes'' two times. At this point it seemed probably right. Finally, on the fifth day, they asked if \textit{medaka} was right, and he answered affirmatively six times out of eight, as shown in Fig. \ref{vectors}B.

\subsubsection{Subject B}
The wh-question, ``How do you like the word communication system?'', was given by his aunt, and his answer was \textit{fuanntei} (unstable). On the first day, he expressed ``U, A, O,'' and she selected \textit{f\textbf{\underline{u}}m\textbf{\underline{ann}}} (dissatisfaction) and \textit{f\textbf{\underline{uann}}} (uneasiness). He answered ``no'' to \textit{fumann} and ``yes'' to \textit{fuann}. On the second day, she confirmed whether ``U, A, O'' was right, and he answered ``yes'' three times out of four. Then, she added \textit{f\textbf{\underline{uann}}tei} to the candidate words, because she thought it was more suitable for her question than \textit{fuann}. She thus gave him a fourfold choice of \textit{fumann}, \textit{fuann}, \textit{fuanntei}, and ``other,'' which actually resulted in \textit{fuanntei}. She then confirmed it twice, and he answered that it was right twice. Finally, on the third day, his aunt confirmed it again, and he answered that it was right three times out of four (for which the value of the statistical measure was 81.4\%). At that point, she accepted the word.

\subsubsection{Subject C}
The wh-question, ``What is your favorite genre for reading?'', was given by his mother. He expressed \textit{rekishi} (history). He used to like listening to his caregivers read books on Roman history, philosophy, and ethics. She thought his taste might have changed and thus asked this question.

On the first day, Subject C expressed ``I, E, O,'' and his mother selected \textit{sh\textbf{\underline{i}}z\textbf{\underline{enn}}} (nature) among 48 candidate words. He denied that word, however, by giving a ``no'' answer. On the second day, she started from the beginning, and he expressed ``O, A, (End).'' Just to make sure, she looked up ``O, A, (End)'' in the dictionary and found \textit{r\textbf{\underline{o}}hm\textbf{\underline{a}}} (Rome) as a candidate word. She asked him if \textit{rohma} was right, and he denied it twice. On the third day, \textit{rohma} was asked again, because he used to like Roman history. He clearly answered ``no,'' however, six times out of eight. Therefore, she started fresh from the beginning, and he expressed ``E, I, A.'' She guessed \textit{\textbf{\underline{ei}}g\textbf{\underline{a}}} (movie) among the candidate words and asked him if it was right, but in vain.

Then, after the third day, his mother noticed that ``E, I, A'' became ``E, I, I'' if the last vowel A was replaced with I, which suggested \textit{r\textbf{\underline{e}}k\textbf{\underline{i}}sh\textbf{\underline{i}}}. Finally, on the fourth day, she asked Subject C if \textit{rekishi} was right, and he answered affirmatively five times out of eight. 

\subsubsection{Subject D}
The wh-question, ``What transportation method do you like to use to return to your hometown?'', was given by her son. She expressed \textit{hikouki} (airplane). Her hometown is far from Tokyo, where she lives, and her son said that air transport was actually the only suitable method among road, rail, air, and sea transport, because the other methods take too much time for her to return to her hometown.

On the first day, Subject D expressed ``I, I, I,'' but her son found no appropriate words in the dictionary. On the second day, she expressed ``O, O, U.'' He then found the appropriate words \textit{h\textbf{\underline{i}}k\textbf{\underline{ou}}ki} and \textit{j\textbf{\underline{i}}d\textbf{\underline{ou}}sha} (automobile) in the dictionary by replacing the first O with I. On the third day, he confirmed whether \textit{hikouki} or \textit{jidousha} was right, and she confirmed \textit{hikouki} seven times out of eight. 

\subsubsection{Subject E}
The wh-question, ``How do you like the word communication system?'', was given by his daughter. On the first day, he expressed ``U, A, I.'' She found 72 candidate words in the dictionary but no appropriate words. On the second day, she confirmed ``U, A, I,'' and he answered ``no'' three times out of four. Then, she started from the beginning and obtained the three vowels ``O, U, (End).'' She found 98 candidate words in the dictionary but no appropriate words again.

Then, on the third day, Subject E expressed ``A, I, A,'' and his daughter found 219 words in the dictionary. Among those words, she selected \textit{\textbf{\underline{a}}r\textbf{\underline{i}}g\textbf{\underline{a}}tai} (thanks) and confirmed it with him. He answered ``yes'' two times and ``no'' two times, to her regret. Finally, on the fourth day she confirmed whether it was right again, and this time, he answered affirmatively five times out of eight.

\section{Discussion}
\subsection{Responses to Expressed Words}

The caregivers' responses to the expressed words differed from person to person. Subject A's word ``killifish'' was a surprise to his wife, because she expected ``dog'' or ``cat.'' He suggested it, however, on the third and fourth days and showed affirmative intention six times out of eight on the fifth day. Then, his wife agreed to accept it. Subject B expressed ``unstable.'' When he affirmed it three times out of four, his aunt wondered if he wanted to stop using the communication aid. Therefore, she asked him if he wanted to keep using it.  He gave affirmative intention three times out of four. Subject C answered ``history,'' which was easily accepted, because his mother had already read him Gibbon's history of Rome. Subject D answered ``airplane.'' Her son selected the transportation question because he knew the appropriate answer, and he was satisfied with it. Subject E answered ``thanks,'' and his family was glad to hear that. Consequently, all the caregivers accepted the words that were expressed by all the subjects.

\subsection{Assist Effect of Caregivers}
To reach a patient's true word, caregiver intervention is essential. Especially, the caregivers' heuristic ability is most helpful. For example, Subject A's wife asked her husband, ``What do you hate most?'', one day after the experiment had finished. He answered ``E, E, E,'' but there were no appropriate candidate words (among only two candidates: \textit{\textbf{\underline{E}}b\textbf{\underline{e}}r\textbf{\underline{e}}suto} (Everest) and \textit{\textbf{\underline{e}}r\textbf{\underline{e}}b\textbf{\underline{e}}htah} (elevator)). She immediately understood, however, that the right word was ``ALS,'' because its pronunciation in Japanese is `` \textit{\textbf{\underline{e}}i} (A), \textit{\textbf{\underline{e}}ru} (L), \textit{\textbf{\underline{e}}su} (S),'' though the first three vowels are ``E, I, E (\textit{\textbf{\underline{eie}}ruesu}).'' When she confirmed ``ALS,'' he showed affirmative intention seven times out of eight. This is a typical case in which the heuristic ability is helpful.

\subsection{Application to Alphabet Letter Matrix}

The structure of English is different from that of Japanese, but our method is also applicable to an alphabet letter matrix. For example, the matrix consists of six rows and six columns. The alphabet letters are grouped according to column numbers: column 1 (A, F, K, P, U, Z), column 2 (B, G, L, Q, V), column 3 (C, H, M, R, W), column 4 (D, I, N, S, X), column 5 (E, J, O, T, Y), and column 6 (space). For example, if a patient chooses the three column numbers, ``4, 1, 5,'' then the candidate words are the following: ``data,'' ``date,''\dots,``suede,'' ``suet''  (131 words) (Hornby, 1974). If the caregiver had asked the patient, ``What genre would you like to listen to?'', then the answer ``4, 1, 5'' could mean ``sports'': S in column 4, P in column 1, and O in column 5.

\subsection{Limitations}
It took 3-5 days for the subjects to express their words (with a 30-minute session each day). It is necessary to improve the online classification accuracy of the ``yes/no'' communication aid to reduce the time needed for patients to express words. In the present experiment, the accuracy of the ``yes/no'' classification during the training period was 70.0\% (SD: 6.5\%) on average over the five subjects (A: 78.6\%; B: 75.0\%; C: 62.5\%; D: 64.3\%; E: 69.5\%). We will thus try to improve the accuracy rate in the near future by developing a method to adjust the parameters for each subject. We expect that the rate will then exceed 80\%.

\section{Conclusion}
Four subjects in almost CLIS and one in CLIS successfully answered wh-questions by using a novel word communication system, and their families were satisfied with the answers. One of the main results was that the proposed system gave one family an unexpected answer, \textit{medaka} (killifish), which meant that the patient and the caregiver could communicate not only in one direction but also mutually.

\vspace{7mm}
\noindent
\textbf{Conflict of Interest Satement}:
Author Siryu Wada was employed by the company Double Research and Development Co., Ltd. The remaining authors declare that the research was conducted in the absence of any commercial or financial relationships that could be construed as a potential conflict of interest.

\vspace{5mm}
\noindent
\textbf{Author Contributions}: 
KO conceived the idea of word expression, mainly conducted the experiments, and wrote the first draft of the manuscript. MN developed the decision method for the communication aid and wrote sections \ref{conf} and \ref{aid}. NT led the whole study. SW was in charge of acquiring funding for the study and conducted a portion of the experiments. All the authors have approved the final manuscript.

\vspace{5mm}
\noindent
\textbf{Acknowledgments}:
We appreciate the participation of all our subjects and our funding sources. We are also grateful to the Japan ALS Association for its cooperation. Part of the study was funded by New Energy and Industrial Technology Development Organization (NEDO, fund number 28F002), which is an independent administrative institution.

\vspace{7mm}
\noindent
\textbf{{\large References}}

\vspace{3mm}

\begin{mybibliography}

{\small
Birbaumer, N., Ghanayim, N., Hinterberger, T., Iversen, I., Kotchoubey, B., K\"{u}bler, A., et al. (1999). A spelling device for the paralysed. Nature, 398, 297-298.

Coyle, S., Ward, T., Markham, C., and McDarby, G. (2004). On the suitability of near-infrared (NIR) systems for next-generation brain-computer interfaces. Physiol. Meas. 25, 815-822. doi:  10.1088/0967-3334/25/4/003

Gallegos-Ayala, G., Furdea, A., Takano, K., Ruf, C. A., Flor, H. and Birbaumer, N. (2014). Brain communication in a completely locked-in patient using bedside near-infrared spectroscopy. Neurology. 82, 1930-1932. doi: 10.1212/WNL.0000000000000449

Halder, S., Takano, K., Ora, H., Onishi, A., Utsumi, K., and Kansaku, K. (2016). An Evaluation of Training with an Auditory P300 Brain-Computer Interface for the Japanese Hiragana Syllabary. Front. Neurosci. 10:446. doi: 10.3389/fnins.2016.00446

Hayashi, H. and Oppenheimer, E.O. (2003), ALS Patients on TPPV; Totally locked-in state, neurologic findings and ethical implementations. Neurology. July 08, 61, 1935-1937. doi: 10.1212/01.WNL.0000069925.02052.1F 

Hong, K., Ghafoor, U., and Khan, M.J. (2020). Brain-machine interfaces using functional near-infrared spectroscopy: a review. Artif. Life Robotics. doi: 10.1007/s10015-020-00592-9
Hornby, A. S. (1974). Oxford Advanced Learner's Dictionary of Current English (Third Edition). London: Oxford University Press.

Miyakoshi, M. Edit. (2010). New Japanese Dictionary for Elementary School (Fourth Edition). Tokyo: Oubunsha.   

Naito, M., Michioka, Y., Ozawa, K., Ito, Y., Kiguchi, M., and Kanazawa, T. (2007). A Communication Means for Totally Locked-in ALS Patients Based on Changes in Cerebral Blood Volume Measured with Near-Infrared Light. IEICE T. Inf. Syst. E90D, 1028-1037. doi: 10.1093/ietisy/e90-d.7.1028

Schalk, G., and Mellinger, J. A. (2010). Practical Guide to Brain-Computer Interfacing with BCI2000. London: Springer. 

Sereshkeh, A.R., Yousefi, R., Wong, A.T., Chau, T. (2019). Online classification of imagined speech using functional near-infrared spectroscopy signals. J. Neural Eng. 16:016005. doi: 10.1088/1741-2552/aae4b9

Simon, N., K\"{a}thner, I., Ruf, C.A., Pasqualotto, E., K\"{u}bler A., and Halder, S. (2015). An auditory multiclass brain-computer interface with natural stimuli: Usability evaluation with healthy participants and a motor impaired end user. Front. Hum. Neurosci. 8:1039. doi: 10.3389/fnhum. 2014.01039

Sitaram, R., Zhang, H., Guan, C., Thulasidas, M., Hoshi, Y., Ishikawa, A., Shimizu, K., and Birbaumer, N. (2007). Temporal classification of multichannel near-infrared spectroscopy
signals of motor imagery for developing a brain-computer interface. NeuroImage 34, 1416-1427. doi: 10.1016/j.neuroimage.2006.11.005

Wolpaw, J. R. and Wolpaw, E.W. (2012). Brain-Computer Interfaces. New York: Oxford University Press.
}

\end{mybibliography}

\end{document}